\documentclass[spmi]{academic} 
\begin{document}
\shortauthor{S. Das Sarma}
\shorttitle{Spintronics: electron spin coherence,...}
\title{Spintronics: 
electron spin coherence, entanglement, and transport}
\author{S. Das Sarma, Jaroslav Fabian, Xuedong Hu, Igor \v{Z}uti\'{c}}
\address{Department of Physics, University of Maryland, College
Park, MD 20742-4111}
%\date{\today}
\maketitle
\begin{abstract}
Prospect of building spintronic devices in which electron spins
store and transport information has attracted strong attention in 
recent years.  Here we present some of our representative theoretical 
results on three fundamental aspects of spintronics: spin coherence, 
spin entanglement, and spin transport.  In particular, we
discuss our detailed quantitative theory
for spin relaxation and coherence in electronic materials, resolving
in the process a long-standing puzzle of why spin relaxation is
extremely fast in Al (compared with other simple metals).  In the
study of spin entanglement, we consider two electrons in a coupled
GaAs double-quantum-dot structure and explore the Hilbert space
of the double dot.  The specific goal is to critically 
assess the quantitative aspects of the proposed 
spin-based quantum dot quantum computer architecture.
Finally, we discuss our theory of spin-polarized transport
across a semiconductor/metal interface.  In particular, we study 
Andreev reflection, which enables us to quantify the degree
of carrier spin polarization and the strength of interfacial
scattering.

\keywords{spintronics, spin coherence, spin relaxation,
spin-hot-spot model, spin entanglement, 
electron exchange, spin transport, Andreev reflection,
spin tunneling} 

\end{abstract}

\section{Introduction}

There has been a great deal of recent interest in the concept of 
spintronics \cite{prinz98} where active control and manipulation
of electron spin in semiconductors and metals provide the basis
of a novel quantum technology.  Possible applications of 
spintronics include high speed magnetic filters, sensors, 
quantum transistors, and spin qubits for quantum computers
\cite{Steane,LD}.  More fundamental
research will, however, be needed before practical
spintronic devices can be demonstrated,
as much remains to be understood about spin coherence, 
spin dynamics, and spin transport.  In this paper we discuss
some of our recent theoretical work on understanding 
spin dynamics in electronic materials.

The existing spintronic architectures \cite{prinz98,gregg97}
and the proposed solid-state quantum computing schemes 
\cite{LD} rely on the relatively long spin coherence
times of conduction electrons. Indeed, in the simplest spintronic 
scheme--the spin injection \cite{johnson85}--electrons with 
a definite spin polarization are supplied into a nonmagnetic metal 
or semiconductor from a ferromagnetic electrode.
The farther (longer) the electrons in the nonmagnetic sample 
carry the spin
coherence, the more useful the device is. Similarly, if an 
electron spin represents
a qubit in a solid-state quantum computer, the longer the 
spin survives, the more
reliably it can store information. The question of how 
spins of mobile
electrons (and holes) lose their spin coherence is thus of 
the utmost importance
for spintronic technology and for solid-state quantum computing.
Unfortunately, the physical picture of spin
decoherence (or relaxation) we now have is far from complete. Most
information comes from
experiments, but experimental data are still scarce 
and often incomplete. The existing theories
seem to provide a broad conceptual framework for understanding
spin decoherence in metals and semiconductors, but the acute 
absence of
realistic calculations for concrete materials makes it 
difficult to validate
these theories. 
The hope is that with more complete experimental and theoretical
understanding we will be able to choose or build materials with 
the longest
decoherence times possible. In this paper we discuss, as an 
example, our
detailed quantitative theory \cite{fabian98,fabian99a,fabian99} 
of spin relaxation and coherence in a simple metal, Al. 

One challenge of spintronics is to study the possibility of using
(electron or nuclear) spins as quantum bits (qubits) in a quantum
computer (QC).  QC has drawn growing attention 
in recent years because it can deliver significant
speed-up over classical computers \cite{Steane} 
for some problems due to inherent
superposition and entanglement of a quantum system.
Various QC architectures have been proposed, including
several solid state models that may possess the important
feature of scalability.  We study  
a QC model \cite{LD} in which
a double quantum dot in the GaAs 
conduction band serves as the basic elementary gate 
for a QC with the electron spins in the dots 
as qubits.  The two-electron exchange coupling provides 
the necessary two-qubit entanglement required for quantum
computation.
Using a molecular orbital approach \cite{McWeeny},
we determine the excitation spectrum of two horizontally 
coupled quantum dots with two confined electrons, and 
study its dependence on an external magnetic field \cite{HD}.  
We particularly focus on the electron exchange coupling 
and double occupation probability, which are two crucial 
parameters for a QC architecture.

Since potential spintronic devices are typically heterojunctions
\cite{johnson85}, it is important to understand how transport 
across the interface between different materials depends on the
degree of carrier spin polarization and the interfacial
transparency.  Significant progress has been made in manipulating
spin dynamics in semiconductors \cite{awschalom97}, including
various methods to create spin-polarized carriers \cite{zs},
such as employing a novel class of ferromagnetic semiconductors
\cite{ohno} and spin injection from a ferromagnet \cite{ham}.
These advances, together 
with  tunable electronic properties (such as
carrier density and Fermi velocity) and well-established 
fabrication techniques, 
provide a compelling reason to study hybrid semiconductor structures
in the context of spintronics \cite{prinz98}. 
In particular, we consider spin-polarized transport in the 
semiconductor/superconductor
hybrid structures, which for low applied
bias is governed by Andreev reflection \cite{and}. 
Our aim is also to motivate study of the interplay between
spin polarization and Andreev reflection in other areas,
such as mesoscopic physics and quantum computing.

\section{Spin coherence in electronic materials}

Spins of conduction electrons decay because of the spin-orbit
interaction and momentum scattering. At low temperatures ($T
\lesssim 20$ K) spin relaxation is caused by impurity scattering and
is temperature independent. At higher temperatures electrons lose
spin coherence by colliding with phonons (phonons can induce a
spin flip because in the presence of a spin-orbit coupling
electronic Bloch states are not spin eigenstates). 
Spin relaxation rate $1/T_1$ increases as temperature
increases, with the growth
becoming linear above the Debye temperature. This mechanism,
discovered by Elliott \cite{elliott54} and Yafet \cite{yafet63},
is the most important spin relaxation mechanism in metals and
semiconductors with inversion symmetry. It gives typical values of
$T_1$ on the nanosecond scale, in agreement with experiment. To
our knowledge the longest $T_1$ in a metal has been reported to be
a microsecond, in a very pure Na sample at low
temperatures \cite{kolbe71}. The situation is much more complicated
in semiconductors \cite{fabian99}. Many interesting semiconductors
like GaAs lack inversion symmetry, so that other mechanisms, in
addition to the Elliott-Yafet one, become
important \cite{fabian99}. These mechanisms operate differently in
different temperature regions, doping, and magnetic, strain, and
confinement fields, so that sorting out the relevant mechanism(s)
for a given material is a tremendous task which is yet to be
carried out. Magnitudes of $T_1$ in semiconductors are also
typically nanoseconds, but recent experimental
studies \cite{awschalom97} in II-VI and III-V systems show that
$T_1$ can be artificially enhanced \cite{fabian99}. It seems,
however, that intrinsic electronic properties may not allow $T_1$
in technologically interesting materials to be longer than a 
microsecond at room temperature.

\begin{figure}
\centering
\includegraphics[width=4in]{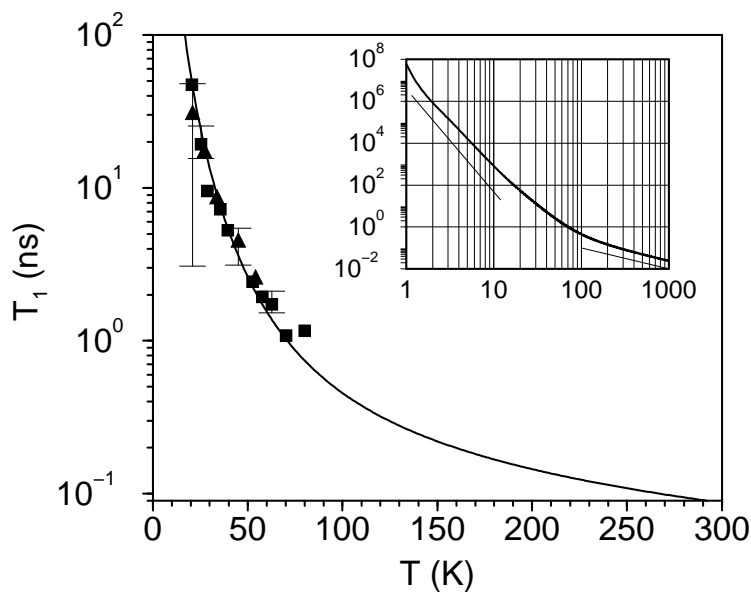}
\caption{Calculated phonon-induced spin relaxation time 
$T_1$ in Al as a function of temperature $T$.
Symbols with error bars are experimental data from Refs.
\cite{johnson85,lubzens76} and the inset is a larger scale 
log-log plot.
There is a nice agreement between the theory and the 
experiment, but the
absence of experimental data at temperatures above 100 K makes
the calculation a prediction which is particularly useful 
for spintronic applications at room temperature.
}
\label{fig:1}
\end{figure}

Spin relaxation is very sensitive to the electronic band
structure. Our {\it ab initio}
calculation \cite{fabian98,fabian99a} of $T_1$ in Al, whose result
is shown in Fig. \ref{fig:1}, shows that band-structure anomalies
like the Fermi surface crossing of a Brillouin zone boundary or an
accidental degeneracy line, can enhance spin relaxation (reduce
$T_1$) by orders of magnitude. Since such anomalies are ubiquitous
in polyvalent metals (the Fermi surfaces of monovalent metals are
well contained within the first Brillouin zone boundary and thus
are free of anomalies), we gave them a special name: spin hot
spots. Whenever an electron jumps in or out (as a result of a
collision) of a spin hot spot, the electron's chance of flipping
spin is greatly enhanced. As a result, $1/T_1$ in Al and other 
polyvalent metals is
much greater than what one would naively expect. This is indeed
what is measured \cite{monod79}: While monovalent alkali and noble
metals have their spin relaxation rates in accordance with simple
estimates based on the Elliott-Yafet theory, polyvalent metals
(only Al, Be, Mg, and Pd have been measured so far) have $1/T_1$
larger than expected by typically one to three orders of
magnitude. Take as an example Al and Na.  They have similar atomic
numbers, so one would expect that their corresponding
spin-orbit couplings would be similar too (as is the case in the
atomic state \cite{yafet63}), giving similar spin relaxation
rates. But the corresponding $T_1$ at the Debye temperatures (150
K for Na and 390 K for Al) are about 20 ns in Na and 0.1 ns in Al!
This huge difference is caused by the presence (absence) of spin
hot spots in polyvalent Al (monovalent Na).

The current fashion for electron spin aside, spin relaxation
in electronic materials is a beautiful and important subject of
its own. The field itself began in the 50's with the advent of
CESR (conduction electron spin resonance), but after some initial
breakthroughs the subject went dormant until the current surge
inspired by spintronics. The experimental focus has been so far on
the simplest elemental metals like the alkali and the noble
metals, and on just a handful of interesting semiconductors. This
is understandable from the point of view of technological
applications, but not quite right from the point of view of
fundamental physical understanding. What is clearly needed is a
catalogue of temperature dependent spin relaxation times for
different metals and semiconductors, a systematic study of the
effects of impurities, alloying, and surfaces and interfaces. 
Theory should have the same goal:
performing realistic calculations of $T_1$ for different metals
within the existing framework laid out by Elliott and Yafet, and
similarly for semiconductors taking into account other spin relaxation
mechanisms as well.

\section{Electron entanglement through exchange interaction 
in a double quantum dot}

\begin{figure}
\centering
\includegraphics[width=5in]{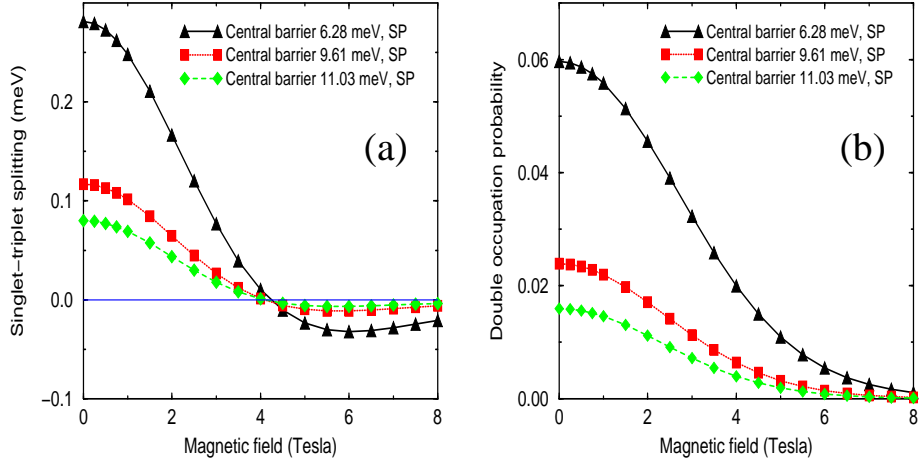}
%\vspace*{0.2in}
\protect\caption[Exchange coupling and double occupation
probability as functions of magnetic field]
{\sloppy{
Calculated magnetic field dependence of
(a) the spin singlet-triplet splitting and (b) double occupation
probability in the singlet state for a two-electron double
quantum dot. The dot size is 30 nm in radius, and the inter-dot 
distance 40 nm.  There is a singlet-triplet crossing at
about 4 Tesla, which is a 
result of competing Coulomb repulsion, exchange, and 
single particle kinetic and potential energies.  DOP 
is relatively high at low magnetic fields, but
is quickly suppressed as magnetic field increases.
}
}
\label{fig2}
\end{figure}

The exchange coupling (the splitting of the lowest
singlet and triplet states) and the
double occupation probability (the probability
that the two electrons occupy 
the same orbital state in one dot) in a double dot
are two important
parameters for the spin-based quantum dot quantum computer (QDQC).  
The exchange coupling between two electrons establishes the
necessary entanglement between spins, 
and determines how fast quantum gates can be.  Quantum
computation has very low tolerance for errors (it requires
an error rate
below $10^{-4}$), so that precise control and small errors are
imperative for a QC.  In a QDQC, individual 
quantum dots are tags which distinguish different qubits.  
If during a gating action two electrons jump onto a single
quantum dot, their original tag information will be lost, 
which will result in an error. 
Thus, in designing a QDQC, double occupation 
probability (DOP) has to be minimized
for the states that belong to the QDQC Hilbert space.
Fig.~\ref{fig2} shows our numerical results on
the magnetic field dependence of (a) the exchange 
coupling with three different 
central barrier heights, and (b) 
the ground state DOP.  The latter clearly
decreases as ${\bf B}$
field increases.  Physically, as ${\bf B}$ increases, the
single-electron atomic wavefunctions are squeezed so that  
the inter-dot wavefunction overlap decreases,
while the ``on-site'' Coulomb repulsion energy for a 
single dot increases.  The ground state DOP
can also be seen in Fig.~\ref{fig2} to decrease
significantly with increasing central barrier strength separating
the two dots, as one would expect.  

As shown in Fig.~\ref{fig2}, 
at zero magnetic field it is difficult to
have both a vanishing DOP for a
small error rate and a sizable exchange coupling for fast
gating, because the exchange coupling and the DOP
have similar dependence on the inter-dot barrier
and inter-dot distance.   
On the other hand, finite magnetic fields 
may provide finite exchange coupling 
for QC operations with small errors.
However, a finite magnetic field will produce a Zeeman
splitting in the triplet state, causing additional
phase shifts.  Therefore, a swap gate \cite{LD} in a QC would
have to include additional single-qubit operations to correct
the effects of these phase shifts \cite{HD}.  This added 
complexity inevitably prolongs the gating time of a two-qubit
operation, which in turn increases the chance of an error due
to spin relaxation.  Another implication
of a finite magnetic field is small exchange coupling---about
an order of magnitude smaller than that at zero field.
This means that the two-qubit operations
will last as long as 10 ns, requiring the spin coherence time
be longer than 10 $\mu$s in a semiconductor quantum dot. 
Whether GaAs or other electronic materials can provide such favorable
environment for electron spins is yet to be determined,
and many other questions need to be answered before practical
spin-based QDQC can be realized.  These questions include,
but are not limited to,
the effects of stray fields, the implication of a chosen geometry,
and the effects of external noise introduced through active control.

\section{Spin transport}

The presence of spin-polarized carriers gives rise to 
both modified charge transport and intrinsic spin transport,
absent in the unpolarized case. Each of these aspects provides
information about the degree of spin polarization which can be
utilized in spintronics.
Here we focus on the transport of spin-polarized
carriers across the semiconductor/metal interface
where the metal is in the superconducting state.
The study of 
semiconductor/superconductor (Sm/S) hybrid structures has several
important ramifications.
Already in the context of spin-unpolarized transport \cite{lamb}, 
it has been
demonstrated \cite{belt} that this configuration can be used to examine
the interfacial transparency which for a Sm/normal metal is typically
limited by a native Schottky barrier.  In the presence of 
spin-polarized carriers, Sm/S structure can also serve to quantify the 
degree of spin polarization of a semiconductor and probe both 
potential and
spin-flip interfacial scattering \cite{zs}.
To understand such sensitivity to spin polarization and 
different types of
interfacial scattering it is important to consider the process 
of Andreev
reflection \cite{and} which governs the low bias transport. In this 
two-particle process, an
incident electron of spin
$\sigma=\uparrow,\downarrow$ on a Sm/S interface is reflected as a
hole belonging to the opposite spin subband, back
to the Sm region while a Cooper pair is transfered to the 
superconductor.
The probability for Andreev reflection at low bias voltage is thus
related to the square of the normal state transmission coefficient and
can have stronger dependence on the junction transparency than 
the ordinary
single particle tunneling.
For spin-polarized carriers, with different populations in two spin
subbands, only a fraction of the incident electrons from a 
majority subband
will have a minority subband partner in order to be Andreev
reflected. 
In the superconducting
state, for an applied voltage smaller than the superconducting gap,
single particle tunneling is not allowed in the S region and the
modification of the
Andreev reflection amplitude by spin polarization or junction 
transparency will be manifested in transport measurements.
%quantities such as the charge
%conductance or current-voltage characteristics of  Sm/S junction.

Prior to work of Ref.~\cite{zs}
the spin-dependent Andreev reflection was addressed only
in the context of ferromagnetic-metal/S junctions \cite{vas}
and calculations were performed assuming 
the equality of the effective masses \cite{zv} in the two regions
across the interface.  Such an assumption is
inadequate for the Sm/S hybrid structures.
We adopt here scattering approach from
Ref.~\cite{zs} and solve the 
Bogoliubov-de Gennes \cite{zs,zv} equations in a ballistic regime. 
At the flat interface between the Sm and S region we model the
interfacial scattering by $Z_\sigma$ and $F$,
orbital and spin-flip scattering strengths, respectively. 
$Z_\uparrow \neq Z_\downarrow$ can describe magnetically
active interface and the effects of spin-filtering.
We represent spin polarization by $X$, the ratio of
spin subband splitting and the Fermi energy.
\begin{figure}
\centering
\includegraphics[width=5in]{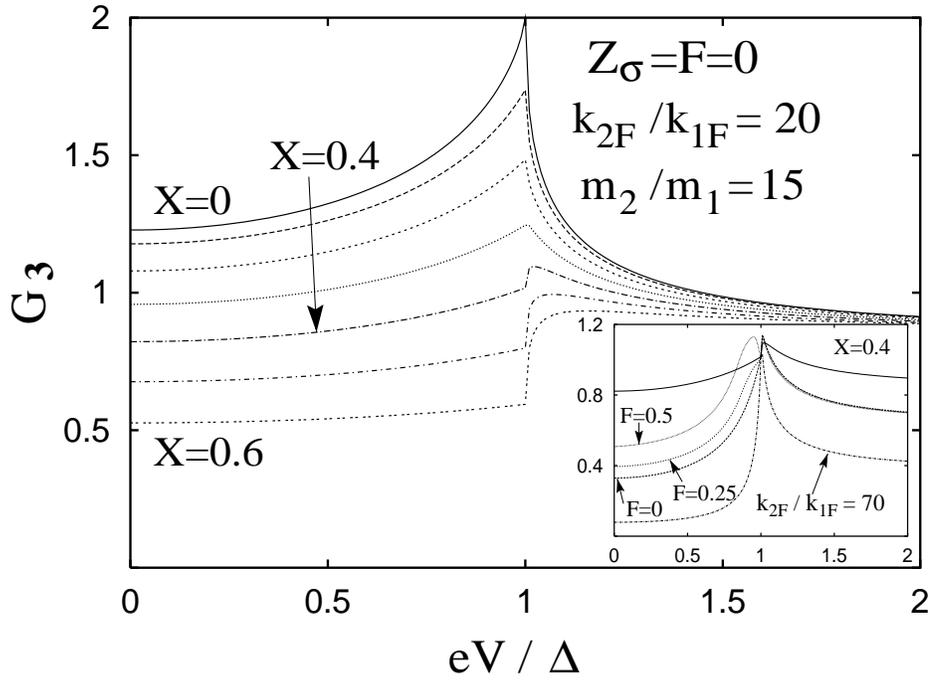}
%\vspace*{0.2in}
\protect\caption[]
{Normalized charge conductance $G_3(eV/\Delta)$.
%
% and spin conductance, $G_{S3}(eV/\Delta)$. In panel (a), 
%
Curves from top to bottom represent $X=0,0.1,0.2,0.3,0.4,0.5,0.6$
at $Z_\sigma=F=0$.  The inset shows $X=0.4$ results.  The 
upper four curves (from top to bottom at zero bias) have $Z_{\sigma}$
and F values of (0,0), (0.5,0.5), (0.5, 0.25), and (0.5,0). 
For the bottom curve, which corresponds to 
$k_{2F}/k_{1F}=70$, the $Z_{\sigma}$ and F values are (0,0).
Here the effective masses and the Fermi wave vectors are 
denoted by $m_i$ and $k_{iF}$, where $i=1,2$ correspond to 
the Sm and S regions respectively.
}
\label{fig3}
\end{figure}
In Fig.~\ref{fig3} we give normalized low temperature results \cite{zs} 
for the three dimensional charge conductance,
$G_3\equiv G_{3\uparrow}+G_{3\downarrow}$, 
as a function of the ratio of bias voltage, $eV$, and the 
superconducting gap, $\Delta$.  Displayed charge conductance,
which is calculated for vanishing interfacial scattering strength,
depends strongly on the spin polarization X.  The inset shows the
effect of orbital and spin-flip scattering at a fixed spin 
polarization of X=0.4.
To study the intrinsic spin transport, it is convenient to define
spin conductance \cite{zs}, $G_{S3} \equiv G_{S3\uparrow} - 
G_{S3\downarrow}$.
For $eV<\Delta$ and any spin polarization 
$G_{S3}=0$, since $G_{S3\uparrow,\downarrow}$
are each proportional to the corresponding
spin component of the quasiparticle current and that 
there is no quasiparticle
tunneling below the superconducting gap. For the 
unpolarized case, $X=0$,
$G_{S3\uparrow} \equiv G_{S3\downarrow}$ and the spin 
conductance vanishes identically. 
For $eV>\Delta$, $G_{S3}$ is a sensitive function of X and 
could be used to determine the degree of the spin polarization.
Experimental studies of  spin-polarized 
Sm/S junctions should 
provide an important test for feasibility of spintronic 
devices based on
hybrid semiconductor structures, as well as stimulate future  
theoretical
studies considering, for example, nonequilibrium processes, realistic 
band structure, and diffusive regime.

\section{Conclusion}

We studied several issues related to spintronics.
First, we
demonstrated the importance of band structure effects 
in spin relaxation
and established that special subtle features 
(spin hot spots) of electronic
structure have profound effects on the magnitude of 
the spin relaxation rate in
electronic materials.  Since spin hot spots can be 
artificially induced, our
work also shows a way of tailoring spin dynamics of 
conduction electrons.
Next, our study of a quantum dot hydrogen 
molecule showed that the goal of having both a reasonable
exchange coupling and a vanishingly small error rate
can only be achieved at finite magnetic fields (4-8 Tesla),
and one has to consider many factors (such as fast gating time,
precise control, low error rate,
etc.) to produce a realistic spin-based quantum computer.
Finally, we demonstrated that the low temperature 
spin-polarized transport
in Sm/S structures may serve as a sensitive and 
quantitative probe 
for determining the degree of spin polarization and 
the strength of interfacial scattering. 
In contrast to the unpolarized case, the junction transparency
can be enhanced with the increase of the Fermi velocity mismatch
in the two regions.

We acknowledge support by US ONR, DARPA, and the Laboratory for
Physical Sciences.


\begin{thebibliography}{99}
%\bibitem[*]{igor} Electronic address: igor@cooperon.umd.edu

\bibitem{prinz98} G. Prinz, Phys. Today {\bf 48}, 58 (1995);
Science {\bf 282}, 1660 (1998).

\bibitem{Steane} 
%D. DiVincenzo, Science {\bf 270}, 255 (1995); 
A. Steane, Rep. Prog. Phys. {\bf 61}, 117 (1998).

\bibitem{LD} D. Loss and D. DiVincenzo, Phys. Rev. A {\bf 57}, 
120 (1998); G. Burkard, D. Loss, and D. DiVincenzo, Phys. Rev. B 
{\bf 59}, 2070 (1999).

%\bibitem{prinz98} G. A. Prinz, Science {\bf 282}, 1660 (1998).

\bibitem{gregg97} J. Gregg. {\it et al.} J. Magn. Magn. Mater. 
{\bf 175}, 1 (1997).

\bibitem{johnson85} M. Johnson and R. H. Silsbee, Phys. Rev. Lett. 
{\bf 55}, 1790 (1985).

\bibitem{fabian98} J. Fabian and S. Das Sarma, Phys. Rev. Lett.
{\bf 81}, 5624 (1998); J. Appl. Phys. {\bf 85}, 5057 (1999).

\bibitem{fabian99a} J. Fabian and S. Das Sarma, Phys. Rev. Lett. 
{\bf 83}, 1211 (1999).

\bibitem{fabian99} J. Fabian and S. Das Sarma, J. Vac. Sc. Technol. B 
{\bf 17}, 1708 (1999).

\bibitem{McWeeny}  
R. McWeeny, {\em Methods of Molecular Quantum Mechanics} 
(Academic Press, San Diego, 1992).

\bibitem{HD} X. Hu and S. Das Sarma, LANL Preprint quant-ph/9911080.
To appear in Phys. Rev. A.

\bibitem{awschalom97} D. D. Awschalom and J. M. Kikkawa,
Physics Today {\bf 52}, 33 (1999).

\bibitem{zs} I. \v{Z}uti\'c and S. Das Sarma,  
Phys. Rev. B {\bf 60}, 16322 (1999).

\bibitem{ohno}  H. Ohno, Science {\bf 281}, 951 (1998).

\bibitem{ham} P.R. Hammar, B.R. Bennett, M.J. Yang and M. Johnson, 
Phys. Rev. Lett. {\bf 83}, 203 (1999).

\bibitem{and} A.F. Andreev, Zh. Eksp. Teor. Fiz. {\bf 46}, 1823 (1964).

\bibitem{elliott54} R. J. Elliott, Phys. Rev. {\bf 96}, 266 (1954).

\bibitem{yafet63} Y. Yafet, in {\it Solid State Physics}, 
edited by F. Seitz and D. Turnbull (Academic, New York, 1963), Vol. 14.

\bibitem{kolbe71} W. Kolbe, Phys. Rev. B {\bf 3}, 320 (1971).

\bibitem{lubzens76} D. Lubzens and S. Schultz, Phys. Rev. Lett. 
{\bf 36}, 1104 (1976).

\bibitem{monod79} P. Monod and F. Beuneu, Phys. Rev. B {\bf 19},
911 (1979).

%\bibitem{Steane} A. Steane, Rep. Prog. Phys. {\bf 61}, 117 (1998).

%\bibitem{Shor} P.W. Shor, in {\it Proceedings of the 35th Annual 
%Symposium on the Foundations of Computer Science}, ed. by 
%S. Goldwasser (IEEE Computer Society, Los Alamitos, CA), p. 124.

%\bibitem{Grover} L.K. Grover, Phys. Rev. Lett. {\bf 79}, 325 (1997).

%\bibitem{Feynman} R.P. Feynman, Int. J. Theor. Phys. {\bf 21},
%467 (1982); Found. Phys. {\bf 16}, 507 (1986).

%\bibitem{DiVincenzo} D. DiVincenzo, in {\em Mesoscopic electron
%transport} ed. by L.L. Sohn, L.P. Kouwenhoven, and G. Sch\"{o}n
%(Kluwer, Dordrecht, 1997).

%\bibitem{johnson85} Johnson  and Silsbee 85

%\bibitem{prinz98} G. Prinz, Phys. Today {\bf 48}, 58 (1995).
%\bibitem{awschalom97}  D.D. Awschalom, J.M. Kikkawa, Phys. 
%Today {\bf 52} (Vol. 6), 33 (1999).

\bibitem{lamb} C.J. Lambert and R. Raimondi, 
J. Phys.: Condens. Matter {\bf 10} 901 (1998);   
C.W.J. Beenakker, Rev. Mod. Phys. {\bf 69}, 731 (1997),
give reviews of unpolarized transport in Sm/S junctions and
address  Andreev reflection in mesoscopic physics.

\bibitem{belt} S. De Franceschi, F. Giazotto, F. Beltram,
L. Sorba, M. Lazzarino, and A. Franciosi, Appl. Phys. Lett. 
{\bf 73}, 3890 (1998).

%\bibitem{ioffe} L.B. Ioffe, V.B. Geshkenbein, M.V. Feigel'man, A.L.
%Fauch\`{e}re and G. Blatter, Nature {\bf 398}, 679 (1999).

\bibitem{vas} V.A. Vas'ko {\it et al}, Appl. Phys. Lett. {\bf 73}, 
844 (1998); R.J. Soulen Jr.{\it et al}, Science {\bf 282}, 85 (1998);
S.K. Upadhyay  {\it et al}, Phys. Rev. Lett.  {\bf 81}, 3247 (1998).

\bibitem{zv} M.J.M. de Jong and C.W.J. Beenakker, 
Phys. Rev. Lett. {\bf 74}, 1657 (1995); 
J.-X. Zhu {\it et al}, Phys. Rev. B {\bf 59}, 9558 (1999);
I. \v{Z}uti\'c and O.T. Valls, Phys. Rev. B {\bf 60}, 6320 (1999); 
{\it ibid.} {\bf 61}, 1555 (2000);
S. Kashiwaya {\it et al}, Phys. Rev. B {\bf 60}, 3572 (1999).

\end{thebibliography}
\end{document}